\documentclass[hidelinks, 10pt]{article}

\usepackage[left=1.0in, right=1.5in, top=1in, bottom=1in, includeheadfoot, nohead]{geometry}
\usepackage[right,mathlines,running]{lineno}
\usepackage{caption}
\usepackage{titling}
\usepackage{authblk}
\usepackage{libertine}
\usepackage{libertinust1math}
\usepackage[T1]{fontenc}
\makeatletter
\renewcommand{\maketitle}{\bgroup\setlength{\parindent}{0pt}
\begin{flushleft}
{\large\@title}
\vspace{0.2cm}\\
\@author\\
\end{flushleft}\egroup}
\makeatother
\usepackage{fancyhdr}
\usepackage{booktabs}

\usepackage{hyperref}
\usepackage[usenames,dvipsnames,svgnames,table]{xcolor}
\usepackage{soul}
\newcommand{\hrefc}[3][blue]{\href{#2}{\color{#1}{#3}}}
\definecolor{myLink}{HTML}{08519c}
\definecolor{mySig}{HTML}{737373}

\definecolor{myBlue}{HTML}{deebf7}

\usepackage{amsthm}
\usepackage{thmtools}
\usepackage[unq]{unique}
\usepackage{wrapfig}
\usepackage{float}
\declaretheoremstyle[
spaceabove=0pt, spacebelow=0pt,
headfont=\normalfont\bfseries,
notefont=\mdseries, notebraces={(}{)},
bodyfont=\normalfont,
postheadspace=0em,
qed=
]{mystyle}
\declaretheorem[style=mystyle, numbered=no, shaded={bgcolor=myBlue, textwidth=0.5\textwidth}]{Significance Statement}

\usepackage[bottom]{footmisc}
\usepackage{sectsty}
\definecolor{mySect}{HTML}{969696}
\sectionfont{\large \color{mySect}}
\subsectionfont{\normalsize \color{mySect}}

\usepackage{amsmath}
\usepackage{bm}
\usepackage{mathtools}
\usepackage{graphicx}
\usepackage{capt-of}
\usepackage[export]{adjustbox}
\usepackage{tabularx}
\usepackage[numbers,sort&compress]{natbib}

\title{Sharp thresholds limit the benefit of defector avoidance in cooperation on networks}

\author[a,b,1]{\hrefc[myLink]{https://orcid.org/0000-0001-9138-3593}{\small Ashkaan K. Fahimipour}}
\author[c]{\hrefc[myLink]{https://orcid.org/0000-0001-9138-3593}{\small Fanqi Zeng}}
\author[c]{\hrefc[myLink]{https://orcid.org/0000-0001-9138-3593}{\small Martin Homer}}
\author[d]{\hrefc[myLink]{https://orcid.org/0000-0001-9138-3593}{\small Arne Traulsen}}
\author[e]{\hrefc[myLink]{https://orcid.org/0000-0001-9138-3593}{\small Simon A. Levin}}
\author[b,f,g,h]{\hrefc[myLink]{https://orcid.org/0000-0001-9138-3593}{\small Thilo~Gross}}

\affil[a]{University of California Santa Cruz,
Institute of Marine Sciences, Santa Cruz, CA, USA}
\affil[b]{University of California Davis,
Department of Computer Science, Davis, CA, USA}
\affil[c]{University of Bristol, 
Department of Engineering Mathematics, Bristol, UK}
\affil[d]{Max-Planck-Institute for Evolutionary Biology, Pl\"{o}n, DE}
\affil[e]{Princeton University,
Department of Ecology and Evolutionary Biology, Princeton, NJ, USA}
\affil[f]{Helmholtz Institute for Functional Marine Biodiversity, Oldenburg, DE}
\affil[g]{Alfred-Wegener-Institute for Marine and Polar Research, Bremerhaven, DE}
\affil[h]{University of Oldenburg,
Institute for Chemistry and Biology of the Marine Environment, Oldenburg, DE}

\begin{document}
\setlength{\bibsep}{0.0pt}
\sloppy
\vspace*{0.5in}
\maketitle
\noindent$^1$Correspondence to:
\texttt{\footnotesize \hrefc[myLink]{mailto://afahimip@ucsc.edu}{afahimip@ucsc.edu}}\\
\vspace{0.25cm} {\color{mySect}\hrule} \vspace{0.25cm}

\noindent
\textcolor{mySig}{
Consider a cooperation game on a spatial network of habitat patches, where players can relocate between patches if they judge the local conditions to be unfavorable. In time, the relocation events may lead to a homogeneous state where all patches harbor the same relative densities of cooperators and defectors or they may lead to self-organized patterns, where some patches become safe havens that maintain an elevated cooperator density. Here we analyze the transition between these states mathematically. We show that safe havens form once a certain threshold in connectivity is crossed. This threshold can be analytically linked to the structure of the patch network and specifically to certain network motifs. Surprisingly, a forgiving defector avoidance strategy may be most favorable for cooperators. Our results demonstrate that the analysis of cooperation games in ecological metacommunity models is mathematically tractable and has the potential to link topics such as macroecological patterns, behavioral evolution, and network topology.
}


\vspace{0.25cm} {\color{mySect}\hrule} \vspace{0.5cm}

\noindent
Cooperation, behavior that leads to benefits for others at a cost to oneself, is widespread across biological systems, ranging from cells cooperating to form organisms, to cooperation among individuals in populations and among micro- and macrobiotic taxa in ecosystems.
In many cases the costs of cooperation are high.
\begin{wrapfigure}[16]{r}{0.5\textwidth}
\vspace{-0.5cm}
\begin{Significance Statement}
{\small
~Cooperators in biological systems often need some advantage to persist in the presence of selfish defectors, but knowing which strategies or behaviors will impact the outcomes of cooperation games remains a challenge. We demonstrate that modeling games as networks of spatially-arranged ecological communities allows one to know whether any strategy will or won't impact the outcomes of any game played on all spatial configurations, without relying on a limited set of examples. By applying the general theory to the so-called snowdrift game, with cooperators who avoid defectors, we show exactly when a defector avoidance strategy can be beneficial, and how the long term outcomes of cooperative behaviors depend on the spatial arrangement of locations where games are played.
}
\end{Significance Statement}
\end{wrapfigure}
Hence, how cooperative behavior persists in a population represents a fundamental question in biology \cite{nowak1992evolutionary, szathmary1995major, doebeli2005models, nowak2006five, von2007theory, pennisi2009origin, butler2019cooperation, muscarella2019species}.
In general, cooperation is most likely to evolve and persist if there are mechanisms that directly or indirectly benefit cooperators' reproductive success. 
Examples include kin selection, punishment of defectors who forgo the cooperative investment, or a direct self-benefit such as in cases of investment into a common good \cite{nowak2006five}.

Among the most general mechanisms that can favor cooperation is the notion of network or spatial \emph{reciprocity} \cite{nowak1992evolutionary, hauert2004spatial, ohtsuki2006simple, li2018punishment}. In classical examples of reciprocity, cooperation creates favorable conditions for other proximal cooperators \cite{nowak2006five}. 
A result is the emergence of cooperative havens, where the rewards generated by mutual cooperation have enriched some physical or topological neighborhoods.
The formation of cooperative neighborhoods in structured populations, where individuals interact with only a limited subset of the population, has traditionally been studied on networks, where each node represents an individual agent and an edge means that the two connected individuals play against each other \cite{nowak1992evolutionary, ifti2004effects, lieberman2005evolutionary, hauert2006spatial, ohtsuki2006simple, langer2008spatial, floria2009social, traulsen2010human, santos2012dynamics, allen2019evolutionary}.
By assuming weak selection and treating space implicitly, the resulting systems can often be analyzed mathematically. Although this framework has become a powerful tool for conceptual understanding, it represents a strong abstraction from real world ecology where interactions, and hence cooperative behaviors, occur often randomly within a location that is itself embedded in a larger spatial context \cite{durrett1994importance, holyoak2005metacommunities, leibold2017metacommunity, gross2020metacom}.
By focusing on spatially-explicit models of cooperation, we gain the opportunity to understand feedbacks between the rules of the game, movement strategies, and long-term persistence of cooperation at larger scales \cite{levin1992problem, durrett1994importance, wakano2009spatial, wakano2011pattern, ohtsuki2006simple, gross2020metacom, hauert2021spatial}.

Here we study a model of cooperation in spatially-structured populations inspired by ecological metacommunities  \cite{gross2020metacom, holyoak2005metacommunities, leibold2017metacommunity, brechtel2018master}, where network nodes --- instead of individuals --- represent habitat patches containing many interacting individuals, and edges mean that two patches are connected by dispersal of those individuals (Fig.~1A). Each patch is a location where games are played, harboring cooperator and defector subpopulations which grow and shrink in time due to internal interactions and movement among locations. Metacommunity models allow one to represent the effects of physical spatial structure directly and explicitly. Moreover, they can be analyzed using master stability functions, which can be used to untangle the impacts of local dynamics and network structure \cite{brechtel2018master, pecora1998master, nakao2010turing}. We use this ability to explore how different movement strategies impact the outcomes of a cooperation game as a function of network structure.

\section*{Results}
We start by illustrating the existence of sharp thresholds in the onset of spatial reciprocity by considering the well-studied snowdrift game \cite{smith1973logic, doebeli2005models} on a spatial network of two patches (Fig.~1A), and showing when a specific dispersal strategy for relocating between patches can lead to the formation of cooperative safe havens for this game.
Most of the assumptions made here will be relaxed in the next section where we present the general theory, which can be extended to any scenario with minor modifications \cite{brechtel2018master}, including to $n$-strategy games \cite{hauert2021spatial}, higher-order interactions \cite{grilli2017higher}, and explicit resource- or object-mediated cooperation \cite{butler2019cooperation, yeakel2020diverse}.

\subsection*{Introductory example}
Consider a spatial network of two linked nodes, where each node $i \in [1, 2]$ is a habitat patch in which organisms live, interact, and reproduce. The link between them represents an avenue of dispersal that individuals occasionally use to disperse to the other patch (Fig.~1A).
Some of the individuals within each patch $i$ are cooperators ($C_i$) who make an investment that creates a shared benefit, whereas other individuals are defectors ($D_i$) who forego this investment.
Within patches, individuals undergo random pairwise encounters defined by a payoff matrix
\begin{linenomath}\begin{equation}
{\Pi} =  \left[ \begin{array}{c c} 
R & S \\ 
T & P \end{array} \right],
\end{equation}\end{linenomath}
that specifies rewards for mutual cooperation $R$ ($C$ encounters another $C$), the sucker's payoff $S$ ($C$ encounters $D$), the temptation to defect $T$ ($D$ encounters $C$), and the punishment for mutual defection $P$ ($D$ encounters another $D$); the payoffs satisfy $P < S < R < T$ to define a snowdrift game \cite{smith1973logic, sugden2004economics}.

We focus first on a straightforward dispersal strategy for cooperators in a spatial setting: attempting to avoid defectors.
Suppose that the defector-avoidance strategy is such that cooperators leave a patch if they have been cheated in multiple consecutive interactions.
Although easy to motivate psychologically, the choice to allow the cooperators to selectively ``walk away'' from a patch remains less explored, as authors are generally hesitant to give cooperators such an ability that confers a direct advantage \cite{aktipis2004know, santos2006cooperation}. 
As we show below, defector-avoidance is not always beneficial for cooperators, often leaving outcomes unchanged. However, when a threshold in network connectivity is crossed, self-organized patterns form where some patches maintain significantly higher or lower cooperator densities indefinitely.

\begin{figure*}[t!]
\centering
\includegraphics[width=0.95\textwidth]{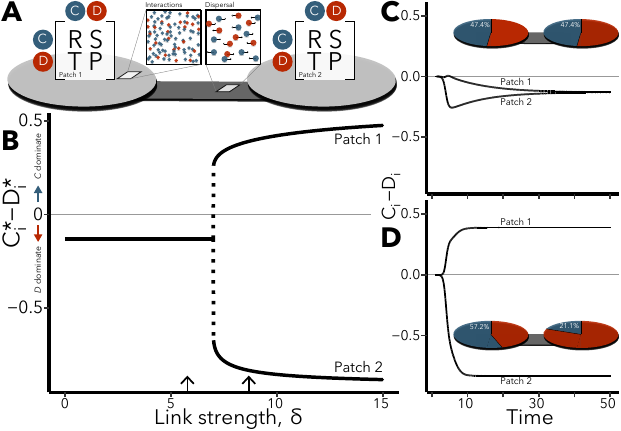}
\captionof{figure}{Emergence of a heterogeneous stationary state on a 2-patch network. \textbf{A.} Schematic of the spatial game, showing local payoff ($\Pi$) relationships among cooperators and defectors occupying the same patch (gray circles) and the dispersal route between them. \textbf{B.} Difference in equilibrium densities of both types in patches 1 and 2 as link strength is varied. Arrows refer to the example time series shown in panels \textbf{C} and \textbf{D}. Initial conditions were uniformly drawn from $[10^{-4}, 10^{-3}]$, and the patch with the largest initial cooperator density is patch 1. \textbf{C.} The homogeneous steady state, with the same equilibrium densities of $C$ and $D$ across locations. The inset network shows the proportions of each type in each patch. \textbf{D.} The same game, but with faster diffusion (larger $\delta$), showing emergence of a heterogeneous steady state with higher cooperator densities in patch 1. Parameters are: $R = 3$, $S = 2$, $T = 5$, $P = 0.2$, $\mu = 1$, $\alpha = 3$.}
\label{fig:fig1}
\end{figure*}

In the model, individuals in both patches are subject to population dynamics of the form
\begin{linenomath}\postdisplaypenalty=0\begin{subequations}
\begin{align}
\dot{C}_i &= G_{{\rm C},i} - M_{{\rm C},i} + \delta \left( E_{{\rm C},j} - E_{{\rm C},i} \right)\\
\dot{D}_i &= G_{{\rm D},i} - M_{{\rm D},i} + \delta \left( E_{{\rm D},j} - E_{{\rm D},i} \right),
\end{align}
\end{subequations}\end{linenomath}
where $G$, $M$ and $E$ are functions of cooperator and defector densities that are described below, and represent the effects of reproduction, mortality and dispersal; and $\delta$ is the link strength of the spatial network.

We assume that the reproduction of individuals is directly proportional to the payoff that they achieve in the game. Using mass-action laws for the encounters, this yields the reproduction rates
\begin{linenomath}\postdisplaypenalty=0\begin{subequations}
\begin{align}
G_{{\rm C},i} &= G_{{\rm C},i}(C_i, D_i) = C_i \frac{R C_i+S D_i}{C_i + D_i} \\ 
G_{{\rm D},i} &= G_{{\rm D},i}(C_i, D_i) = D_i \frac{T C_i+P D_i}{C_i + D_i},
\end{align}
\end{subequations}\end{linenomath}
where, following \cite{durrett1994importance}, intrinsic growth and encounter rates are accommodated as part of $R$, $S$, $T$, and $P$. We assume density-dependent mortality, resulting in
\begin{linenomath}\postdisplaypenalty=0\begin{subequations}
\begin{align}
M_{{\rm C},i} &= M_{{\rm C},i}(C_i, D_i) = \mu C_i (C_i+D_i) \\
M_{{\rm D},i} &= M_{{\rm D},i}(C_i, D_i) = \mu D_i (C_i+D_i),
\end{align}
\end{subequations}\end{linenomath}
where $\mu$ is a rate constant. Finally, the effects of dispersal are
\begin{linenomath}\postdisplaypenalty=0\begin{subequations}
\begin{align}
E_{{\rm C},i} &= E_{{\rm C},i}(C_i, D_i) = C_i Z_{{\rm C},i} \\
E_{{\rm D},i} &= E_{{\rm D},i}(C_i, D_i) = D_i Z_{{\rm D},i},
\end{align}
\end{subequations}\end{linenomath}
where $Z$ is the \emph{per capita} rate at which individuals leave a habitat. 

Here, we consider a situation where defectors disperse at a constant rate $Z_{{\rm D},i} = 1$, whereas cooperators leave if they have been cheated $\alpha$ times in a row (see \textsl{Materials \& Methods})
\begin{linenomath}\begin{equation}\label{eq:gen_m_diff}
Z_{{\rm C},i} = Z_{{\rm C},i}(C_i, D_i) = \left( \frac{D_i}{C_i+D_i} \right)^\alpha.  
\end{equation}\end{linenomath}
Exploring the model numerically for $\alpha=3$ (Fig.~\ref{fig:fig1}), we find that at low link strengths $\delta$ (\emph{i.e.}~low diffusion rates) the system approaches a homogeneous stable state, where each patch harbors the same relative densities of cooperators and defectors (Figs.~\ref{fig:fig1}B,~\ref{fig:fig1}C; equilibrium densities are denoted by $C_i^\star$, $D_i^\star$). In this example, defectors are the most abundant type in all habitat patches, $C_i^\star - D_i^\star < 0$ for all $i$ (Fig.~\ref{fig:fig1}B). When the link strength is increased beyond a critical point, then the homogeneous state becomes unstable to perturbations and the system undergoes a bifurcation and instead approaches a heterogeneous state (Figs.~\ref{fig:fig1}B,~\ref{fig:fig1}D), where the cooperators constitute a majority in one patch while they largely abandon the other.

\subsection*{General theory}
We now describe a general theory for the stability of homogeneous states in a broad class of games on arbitrary patch networks, using a master stability function approach \cite{brechtel2018master, pecora1998master}. Consider a game with the following properties: 
\emph{i}) the interaction dynamics within a patch can be faithfully modeled by a system of differential equations, and 
\emph{ii}) if played on a single patch the system will approach a stationary state. Now consider this game on a network of patches, where 
\emph{iii}) patches are of identical quality, 
\emph{iv}) links are bidirectional and lossless, and \emph{v}) the emigration rate from a patch is proportional to the number of links.
These conditions do not exclude very high-dimensional systems, strong nonlinearities, strong selection in the evolutionary dynamics, or complex decision rules (\emph{e.g.}~cross-diffusion, adaptive dispersal \cite{gross2020metacom}).

Under the conditions above, at least one steady state exists where the communities in each patch are identical (\emph{e.g.} Fig.~\ref{fig:fig1}C); we call these states \emph{homogeneous}. In homogeneous states, community compositions are independent of spatial network topology and can be found, even for very large networks, by analyzing a patch in isolation (see \emph{Materials \& Methods}). However, the stability of homogeneous states is sensitive to network topologies and thus stable homogeneous behavior may be possible on some patch networks, while instability may lead to heterogeneous behaviors emerging in others \cite{brechtel2018master, segel1976application, levin1974dispersion, levin1985pattern}.

The stability of homogeneous states can be computed from local linearizations of the dynamics, captured by the Jacobian matrix $\bf J$. 
For a model with $N$ heritable types or player strategies per patch and $M$ patches, $\bf J$ has the dimension $NM\times NM$. However, the Jacobian is not an unstructured matrix, but instead intricately reflects the structure of the system, which we can make explicit by writing
\begin{linenomath}\begin{equation}
\label{JacSplit}
{\bf J}= {\bf I}\otimes{\bf P} - {\bf L} \otimes {\bf C}
\end{equation}\end{linenomath}
where $\bf I$ is an $N\times N$ identity matrix, $\bf P$ is the Jacobian matrix for the game played on an isolated patch, the coupling-matrix $\bf C$ is a Jacobian-like matrix that consists of partial derivatives of the emigration rates from one patch with respect to population sizes in that patch, $\bf L$ is the weighted Laplacian matrix of the patch network, and $\otimes$ is a Kronecker product \cite{othmer1971instability, segel1976application, brechtel2018master} (see \emph{Materials \& Methods} for details on these matrices).

A stationary state is stable if all eigenvalues of the Jacobian matrix, ${\rm Ev}({\bf J})$, have negative real parts. Using Eq.~(\ref{JacSplit}) these eigenvalues can be computed as 
\begin{linenomath}\begin{equation}
\label{MSF}
{\rm Ev}({\bf J})=\bigcup_{m=1}^M {\rm Ev}({\bf P}-\kappa_m {\bf C})
\end{equation}\end{linenomath}
where $\rm Ev$ returns the set of eigenvalues of a matrix and $\kappa_m$ is the $m$th eigenvalue of $\bf L$ \cite{brechtel2018master} (\emph{Materials \& Methods}). An attractive feature of this approach is that it separates the impact of spatial network structure encoded in $\bf L$ from the effect of local dynamics. Specifically, it shows that the spatial network structure can affect the stability of the system only via the Laplacian eigenvalues, $\kappa$.

An alternative interpretation of Eq.~\ref{MSF} is to view $\kappa$ as an unknown, real-valued parameter, and define a master stability function that captures the general relationship between the structure of all patch networks and pattern-forming instabilities.
The master stability function can be defined as
\begin{linenomath}\begin{equation}
\label{eq:msfequation}
S(\kappa) = {\rm Ev}_{\max}\left({\bf P}-\kappa {\bf C}\right).
\end{equation}\end{linenomath}
where ${\rm Ev}_{\max}$ denotes the eigenvalue with the largest real part.
If a particular value of $\kappa$ leads to a positive $S$, $S(\kappa) > 0$, then we can say that any network with that Laplacian eigenvalue $\kappa$ will be susceptible to pattern-forming instabilities for a particular game (${\bf P}$) and movement strategy (${\bf C}$). Because the effect of space is thus encapsulated in the Laplacian eigenvalues, the remaining eigenvalue problem in Eq.~(\ref{eq:msfequation}) is easier since the relevant matrix has the size $N\times N$, even for very large spatial networks.

\begin{figure}[t!]
\centering
\includegraphics[width=0.5\textwidth]{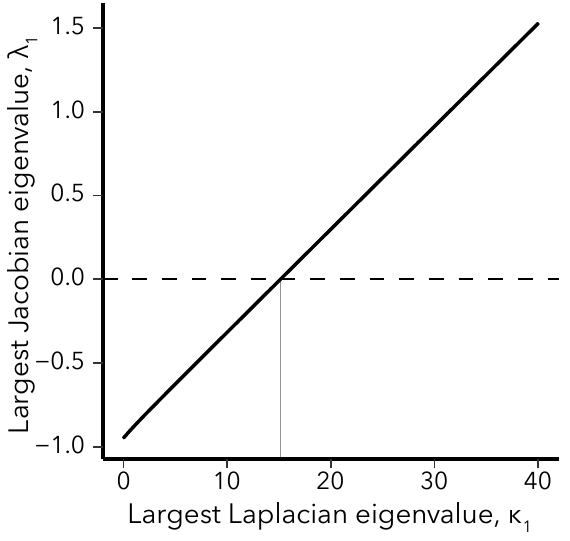}
\captionof{figure}{The appearance of heterogeneous stationary states on abitrary networks. \textbf{A.} Master stability function (Eq.~\ref{eq:msfequation}) of the example snowdrift game. A vertical grey line marks $\kappa_{\rm crit}$ for this game, above which spatial patterns emerge.}
\label{fig:fig2}
\end{figure}
To illustrate the master stability function let us return to the game from the introductory example, which we now consider on arbitrary networks described by a weighted adjacency matrix ${\bf A}$, such that link weight between node $i$ and $j$ is $A_{ij}$. In this more general case the game is described by the following equations
\begin{linenomath}\begin{eqnarray}
\dot{C}_i &=& G_{{\rm C},i} - M_{{\rm C},i} -w_i E_{{\rm C},i}+\sum_j A_{ij} E_{{\rm C},i} \\
\dot{D}_i &=& G_{{\rm D},i} - M_{{\rm D},i} -w_i E_{{\rm D},i}-\sum_j A_{ij} E_{{\rm D},i}.
\end{eqnarray}\end{linenomath}
where $w_i=\sum_j A_{ij}$ is the weighted degree of $i$. 

Using the same parameters as before (Fig.~1), the nonspatial Jacobian ${\bf P}$ and the coupling matrix ${\bf C}$ calculated at equilibrium (see \emph{Materials \& Methods}) are
\begin{linenomath}\begin{equation}
\label{eq:numerical_eg_matrices}
{\bf P} =  \left[ \begin{array}{c c} -0.92 & -1.4 \\ 0.028 & -2.5\end{array}\right] \quad\quad\quad
{\bf C} =  \left[ \begin{array}{c c} -0.06 & 0.19 \\ 0 & 1\end{array}\right] \quad
\end{equation}\end{linenomath}
This leads to the master stability function (Fig.~\ref{fig:fig2})
\begin{linenomath}\begin{equation}
S(\kappa) = \frac{1}{2} \left(\sqrt{1.13 \kappa ^2 + 3.33 \kappa + 2.33} - 0.94 \kappa - 3.42\right)
\end{equation}\end{linenomath}
We can see that $S>0$ on any network that has a Laplacian eigenvalue $\kappa > 15.13$; we refer to this as the critical $\kappa$, or $\kappa_{\rm crit}$, which is specific to the game, but independent of the network structure on which the game is played.

For example a pair of nodes connected by a single link of weight $\delta$ has a leading eigenvalue of $\kappa_1=2\delta$. This shows that the homogeneous state in our example game must become unstable on such an isolated link if $\delta > 7.56$, which explains our previous observations (Fig.~\ref{fig:fig1}B). Together, the results from this section illustrate that the master stability function approach can be used to disentangle the impacts of game parameters from the impact of the topological structure of the underlying network.

\subsection*{Impact of network motifs}
There is a wealth of mathematical knowledge that links Laplacian eigenvalues to specific network properties. Because the Laplacian is symmetric, it must have an eigenvalue $\kappa$ that is greater or equal to the largest eigenvalue in any subgraph of the network \cite{anderson1985eigenvalues}. Hence if any motif in the network has an eigenvalue $\kappa>\kappa_{\rm crit}$, the whole network must also have such an eigenvalue and the homogeneous state must be unstable.

The subgraph rule allows us to extend our results on isolated links in the network to any link in the network. We can for instance say that the homogeneous state is unstable if there is any link of strength $\delta>\kappa_{\rm crit}/2$.   
Similar criteria can be constructed for any conceivable motif. For example a node that is connected to $n$ other nodes via links with a strength of least $\delta$ has an eigenvalue $\kappa\geq (n+1)\delta$. This shows that the homogeneous state in our example game is definitely unstable if there is a node that has at least 15 links of strength 1 or more. 
It is also possible to derive sufficient criteria for stability of the homogeneous state. For example Gershgorin's theorem implies that any Laplacian eigenvalue obeys $\kappa\leq 2k_{\rm max}$ where $k_{\rm max}=\max_j \sum_i A_{ij} $ is the maximum weighted node degree in the network \cite{anderson1985eigenvalues}. Thus in the example game the homogeneous state is guaranteed to be stable if $k_{\rm max}<\kappa_{\rm crit}/2$ (\emph{e.g.}, Fig.~1B).

The examples in the present section illustrate that one can derive topological stability criteria that link dynamical transitions to features of the network, such as the presence or absence of certain network motifs. Additional rules for specific kinds of networks (\emph{e.g.}, regular graphs, lattices) can also be derived \cite{allen2019evolutionary, nakamaru1997evolution}. Such criteria are particularly easy to formulate for games that are characterized by rising master stability functions, including our example system. As a next step we explore the conditions under which a game has this property. 

\subsection*{The value of forgiving dispersal strategies}
Diffusion generally has an equalizing effect that favors homogeneous outcomes \cite{holt2002food, brechtel2018master, gross2020metacom}. Nonlinear mortality, which is needed in the model to avoid boundless growth, constitutes a further dampening force that drives the system to homogeneity. 
It can be shown that strong nonlinearities in the movement behaviour are necessary to overcome this dampening and allow pattern formation.
In our example game, the rule that a cooperator leaves on average, when cheated $\alpha$ times in a row, leads to factors $\left[ D/(C+D) \right]^\alpha$ (\textsl{Materials \& Methods}). One can quickly verify that an overzealous dispersal strategy, where one leaves after being cheated for the first time, isn't nonlinear enough to destabilize the homogeneous state. Thus, in a world where everybody is eager to emigrate to avoid defectors, emigration is useless as the conditions would become identical in all nodes. By contrast, a more forgiving strategy where agents only disperse after having been cheated 10 times in a row, leads to very nonlinear dispersal functions, that are likely to destabilize the homogeneous state. Thus, a forgiving dispersal strategy, may be rewarded indirectly by the formation of safe havens for cooperation.

\begin{figure}[t!]
\centering
\includegraphics[width=0.7\textwidth]{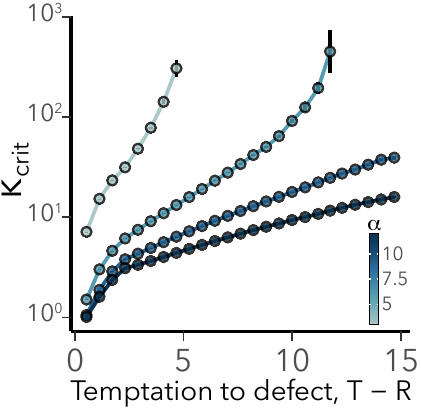}
\captionof{figure}{Correlation between $\log_{10} \kappa_{\rm crit}$ and key parameters: normalized temptation, $T - R$ and the tolerance for consecutive defector encounters, $\alpha$. Points and error bars show mean $\pm$ S.E.M, which are too small to see for most values. Parameter ranges are given in Table~\ref{tab:ranges}.}
\label{fig:fig3}
\end{figure}

We tested the counter-intuitive benefit of forgiving dispersal strategies in numerical experiments, where we considered a large number ($2 \times 10^7$) of feasible steady states 
in systems with different parameter values (see Table~\ref{tab:ranges}, \emph{Materials \& Methods}). We focus only on systems where the homogeneous state is stable at sufficiently low values of coupling, and then ask how much the coupling strength has to be increased to trigger instability. The results show that more forgiving strategies, where cooperators endure more contiguous interactions with defectors before emigrating on average, consistently leads to instability at lower coupling strength, even as the benefits to defection increase (Fig.~\ref{fig:fig3}). This provides further evidence that forgiving dispersal strategies favor the formation of spatially heterogeneous states. 

\subsection*{Locations of safe havens for cooperation}
So far we have shown that defector avoidance has no impact on the outcomes of the game unless certain thresholds are crossed, which in turn can be linked to topological features of the underlying network.
We now use simulations to explore the behavior beyond this threshold. For illustration we consider 100-node random geometric graphs, which provide a reasonable approximation for real networks of habitats and the dispersal connections between them \cite{gross2020metacom}. The coupling strength $\kappa$ is chosen for each simulation such that it exceeds the threshold value (Eq.~\ref{eq:msfequation}) at which the homogeneous state becomes unstable.

Visual inspection of simulated metacommunities quickly reveals that some of the nodes become cooperator dominated. Moreover, these safe havens of cooperation seem to occupy locally well-connected nodes, but not the most highly-connected nodes in the whole network (Fig.~\ref{fig:fig4}A).

In network science the number of connections (degree centrality) is a relatively crude notion of the topological importance of a node in the network. A more sophisticated metric is provided by adjacency-based eigenvector centrality \cite{bonacich1972factoring}, which is loosely related to Google's PageRank \cite{ilprints422} algorithm.
Analyzing an ensemble of 1000 network simulations (parameters as in Fig.~\ref{fig:fig1}) reveals that the nodes of lowest eigenvector centrality become defector dominated, whereas better connected nodes with a higher centrality can sustain a majority of cooperators. However, the most-central nodes in each network are a toss-up, containing almost equal populations of defectors and cooperators (Fig.~\ref{fig:fig4}B). Hence, at least in our example game, the locations where cooperative safe havens form are highly connected nodes, but not the most-highly connected nodes, and this pattern is seen consistently across distinct spatial networks.

\begin{table}[b]
\centering
\begin{tabular}{lrr}
Parameter & Interpretation & Value \\
\midrule
R & Reward from mutual cooperation & $U(0.25, 2.5)$ \\
S & $C$ reward when encountering $D$ & $R - z R$, $z \sim U(10^{-2}, 1)$ \\
T & $D$ award when encountering $C$ & $z R$, $z \sim U(2, 7)$ \\
P & Punishment from mutual defection & $S - z S$, $z \sim U(10^{-2}, 1)$ \\
$\mu$ & \emph{per capita} mortality rate & $U(0, 1)$ \\
\bottomrule
\end{tabular}
\caption{Parameter values for numerical experiments.}
\label{tab:ranges}
\end{table}


\subsection*{Shortwave instabilities and other dispersal strategies}
A distinctive feature of the example game is that it is unstable for sufficiently high values of $\kappa$. Drawing on an analogy with pattern formation in continuous space we call this behavior shortwave instability.   

As pointed out in \cite{segel1976application,nakao2010turing,brechtel2018master} there is a deep analogy between the master stability function on networks and the Turing instability in partial differential equations (PDEs). The master stability function equation becomes identical to Turing's seminal approach if we replace the negative network Laplacian $-{\bf L}$ with the Laplace operator in continuous space. The eigenvalue $\kappa$ can then be interpreted as a wave number. A rising master stability function shows that the instability is most pronounced at arbitrarily high wave numbers, i.e.~arbitrarily short waves, which would be unphysical in PDE systems, but is meaningful in a network. 

To explore when shortwave instabilities occur, consider that, except for some pathological cases, we can assume that as the weighted degree of at least one node in the network approaches infinity
\begin{linenomath}\begin{equation}
\label{eq:limit}
    \lim_{\kappa \to \infty} {\rm Ev}_{\max}({\bf P}-\kappa {\bf C}) = -\kappa {\rm Ev}_{\max}( {\bf C}),
\end{equation}\end{linenomath}
as $\bf P$ becomes negligible in comparison to $\kappa {\bf C}$. This shows that the shortwave instability occurs when the dispersal strategy is such that $\bf C$ has a negative eigenvalue. For games with two types ($C$ and $D$) the coupling matrix has the form. 
\begin{linenomath}\begin{equation}
    {\bf C}=\left(\begin{array}{c c} 
    \partial_C E_{{\rm C}} & \partial_D E_{{\rm C}} \\
    \partial_C E_{{\rm D}} & \partial_D E_{{\rm D}} \\
    \end{array} \right)
\end{equation}\end{linenomath}
At least one eigenvalue with negative real part exists if either
\begin{linenomath}\begin{equation}
    0>\lambda_1+\lambda_2={\rm Tr}({\bf C})=\partial_C E_{{\rm C}}+\partial_D E_{{\rm D}} 
\end{equation}\end{linenomath}
or
\begin{linenomath}\begin{equation}\label{eq::dispcond}
    0>\lambda_1\lambda_2=|{\bf C}| =  \partial_C E_{{\rm C}}\partial_D E_{{\rm D}}-\partial_D E_{{\rm C}}\partial_C E_{{\rm D}} 
\end{equation}\end{linenomath}
One can think of the two terms in the first condition as the degree to which cooperators promote the emigration of cooperators ($\partial_C E_{{\rm C}}$) and vice versa for defectors. Hence the first condition is met if cooperators suppress the emigration of cooperators strongly enough to overcome the effect of defectors promoting their own emigration.  
Assuming that presence of defectors promotes the emigration of both cooperators and defectors we can write the second condition as 
$\frac{ \partial_C E_{{\rm C}}}{\partial_D E_{{\rm C}}}  < \frac{\partial_C E_{{\rm D}}}{\partial_D E_{{\rm D}}}$.
The fraction on the right-hand side can be assumed to be negative or zero because the presence of cooperators should reduce defector emigration or leave it unchanged in reasonable models. By contrast the left hand side can be positive as $\partial_C E_{{\rm C}}$ can either be negative, due to retention of cooperators in a cooperative environment, or positive due to the mass-action effect leading to a positive scaling of cooperator emigration with cooperator numbers. However, the condition can again be satisfied if cooperators suppress their own emigration strongly enough.

Summarizing these results, we can say that shortwave instabilies are primarily expected in those systems where cooperators strongly (nonlinearly) increase the retention of other cooperators in their patch. The defector avoidance rule analyzed in this paper is a special case of this general condition.

\section*{Discussion}
We showed that ecologically motivated models of cooperation games on networks can be studied mathematically. In particular, the master-stability function approach from synchronization provides a powerful tool to explore when a particular game will lead to heterogeneous states where spatial reciprocity becomes possible and safe havens for cooperation can be formed.

\begin{figure*}[t!]
\centering
\includegraphics[width=\textwidth]{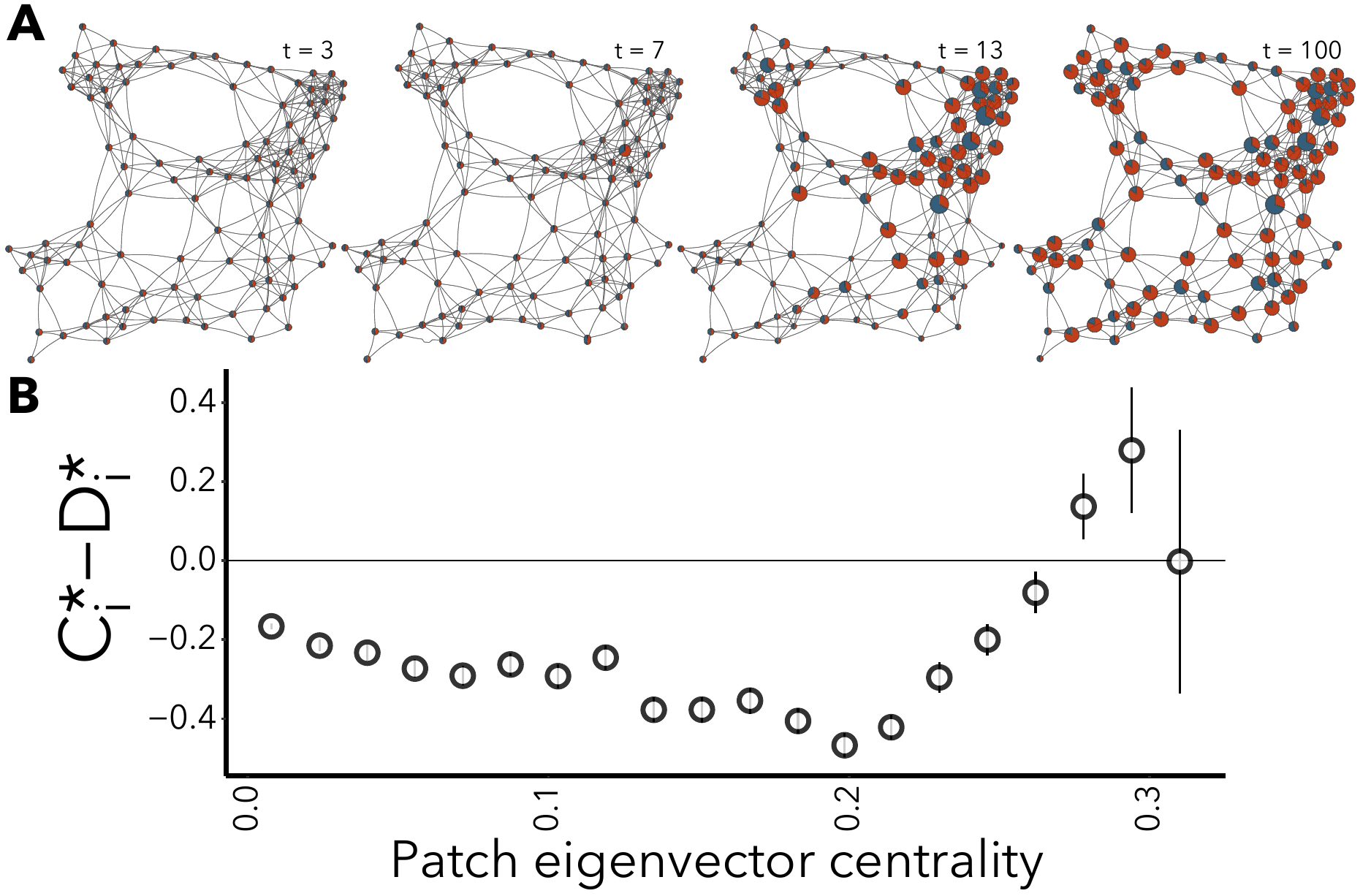}
\captionof{figure}{\textbf{A.} Snapshots of dynamics on an example network with $\kappa_{1} > \kappa_{\rm crit}$ (grey line in Fig.~2). Nodes show the proportion of cooperators (blue) and defectors (red) respectively; node radius is proportional to $|C^\star_{i} - D^\star_{i}|$. Parameters as in Fig.~\ref{fig:fig1}. \textbf{B.} Simulations on 1000 random geometric graphs, showing the association between relative cooperator densities at equilibrium ($C_i^\star - D_i^\star$) and patch eigenvector centrality (bin means $\pm$ S.E.M.).}
\label{fig:fig4}
\end{figure*}

A focus of prior work has been on understanding the evolution and persistence of cooperation in structured populations, where individuals interact through pairwise encounters that constitute a network \cite{nowak1992evolutionary, hauert2004spatial, lieberman2005evolutionary, hauert2006spatial, langer2008spatial, roca2009evolutionary}, or via diffusive public goods \cite{butler2019cooperation, butler2018stability, muscarella2019species}, sometimes on a featureless, continuous spatial plane \cite{wakano2009spatial, wakano2011pattern}.
In this study, we build on this work by studying populations that are structured in a different way, namely as patchy communities where interactions occur randomly within patches and movement among patches in a spatial landscape define the network links (Fig.~1A). 
We find that cooperation can thrive in metacommunities, but that it is mediated by the ability of cooperators and defectors to move between patches. Our findings indicate that the specific movement strategies deployed by both cooperators and defectors are an important factor in the maintenance of cooperation in spatially-structured populations (\textsl{e.g.} Eq.~\ref{eq::dispcond}).
Future work could explore this result in metacommunities of heterogeneous habitat patches \cite{brechtel2019far, anderson2021body} or with lossy links; with individuals who follow adaptive dispersal strategies \cite{abrams2007habitat}; or in systems with multiple interaction types \cite{wilson1997cooperation, pilosof2017multilayer, yamauchi2018spatial, yeakel2020diverse}.
Extending our results to non-stationary dynamics will also be useful for understanding more complicated games that show different types of pattern-forming instabilities (\textsl{Materials \& Methods}).

The work presented here revealed two main findings which some readers may find counter-intuitive: First, allowing cooperators to emigrate selectively \cite{gross2020metacom, aktipis2004know, santos2006cooperation}, in response to defector density, does not always confer a direct benefit to the cooperators. Defector 
avoidance can only result in an increased payoff for cooperators if it is sufficiently strong to overcome a sharply defined threshold, where the system leaves the homogeneous state. The master stability function approach allows us not only to compute this threshold precisely, but to disentangle the effects of the game and the underlying network topology. This opens up a promising angle for future investigations on the impact of specific scenarios and specific network motifs. 

Our second major finding concerns the role of forgiving dispersal strategies in triggering shortwave instabilities. The shortwave instability is a genuine network effect that would not be observed in continuous geometries. Based on our findings we expect this instability to occur particularly if the cooperators respond strongly nonlinearly to cooperation levels. Namely, the instability may be triggered by forgiving dispersal strategies where the cooperator only leaves a patch after being cheated several times in consecutive games. Therefore, forgiving dispersal strategies may be far more beneficial than stricter responses. In many scenarios only the forgiving strategy will induce the heterogeneity in the system that ultimately creates safe havens for cooperation, whereas a stricter more immediate response to defection will result in maladaptive dispersal in a system of identical patches.

One possible criticism may be that even in the heterogeneous state, cooperation doesn't become widespread but mostly remains confined to some nodes which typically occupy central (but not most-central) positions in the network. We nevertheless believe that the formation of such hubs for cooperation can be an important stepping stone in the evolution of higher forms of cooperation and social complexity. Beyond the scope of the class of models explored here, the formation of local cooperation hubs may enable secondary processes, such as the formation of social norms and governance structures, which once established can help promote cooperative behavior in the rest of the network.

\section*{Materials \& Methods}
\subsection*{Patch steady states and stability}
The class of systems considered here have homogeneous stationary states where all nodes approach the same state regardless of the topology of the underlying spatial network (Figs.~1B \& 1C). In these states the net biomass flows in and out of each patch must be equal, such that neither dispersal (selective or otherwise) nor network topology can affect population densities. In any homogeneous state, the densities of cooperators and defectors in any patch are therefore identical to densities in the nonspatial case \cite{brechtel2018master} described by
\begin{eqnarray}
\dot{C} &=& C \frac{RC + SD}{C + D} - \mu C (C + D)\label{eq:Cpatch} \\
\dot{D} &=& D \frac{TC + PD}{C + D} - \mu D (C + D) \label{eq:Dpatch}.
\end{eqnarray}

Setting time derivatives to $0$ in Eqs.~\ref{eq:Cpatch} and \ref{eq:Dpatch}, we find that the system has the three following homogeneous steady states that describe biomass densities across $i$ identical patches: (i) only cooperators persist, with $C_i^\star = R / \mu$ and $D_i^\star = 0$, (ii) only defectors persist, with $C_i^\star = 0$ and $D_i^\star = P / \mu$, and (iii)
a coexistence state of cooperators and defectors, with 
\begin{eqnarray}
C_i^\star = \frac{(P-S) (P R-S T)}{\mu (P+R-S-T)^2},\\
D_i^\star = \frac{ (R-T)(PR-ST)}{\mu (P+R-S-T)^2}.
\end{eqnarray}

The coexistence state is only biologically feasible if $C^\star_i>0$ and $D^\star_i>0$, which places conditions on the relative payoffs each type of player can receive from interactions. This holds under two sets of conditions. The first occurs when $P > S$ and $R > T$. In these cases, the payoff from an interaction with a defector is larger for defectors while the payoff from an interaction with a cooperator is larger for cooperators. Alternatively, positivity occurs when $P < S$ and $R < T$. This case includes the classical ``snowdrift'' game \cite{sugden2004economics}: a cooperator meeting a defector pays the entire cost but still experiences the benefits, while a defector encountering another defector results in no benefit to either ($P < S$). Meanwhile, a cooperator meeting another cooperator invests a fraction of the cost, while a defector meeting a cooperator gets the benefit for free ($R < T$).

The within-patch Jacobian matrix ${\bf P}$ in the coexistence steady state is
\begin{equation}
\begin{split}
{\bf P} &= \left(\begin{array}{c c} 
    \partial_C \dot{C} & \partial_D \dot{C} \\
    \partial_C \dot{D} & \partial_D \dot{D} \\
    \end{array} \right)\\
&= \left(
\begin{array}{cc}
 -\frac{(P-S) (R (P-R+S)+T (R-2 S))}{(P+R-S-T)^2} & \frac{(P-S) (P (S-2 R)+S (R-S+T))}{(P+R-S-T)^2} \\
 \frac{(R-T) (P (T-2 R)+T (R+S-T))}{(P+R-S-T)^2} & -\frac{(R-T) (P (R+S+T) - 2 S T - P^2)}{(P+R-S-T)^2} \\
\end{array}
\right)
\end{split}
\end{equation}
which has eigenvalues
\begin{equation}
\lambda_1=\frac{(P - S) (R - T)}{P + R - S - T},\quad \lambda_2=\frac{S T - P R}{P + R - S - T}.
\end{equation}
Thus when $P > S$ and $R > T$, $\lambda_2 > 0$ and the system is always unstable. By contrast, if $P < S$ and $R < T$, $\lambda_2 < 0$ and so the state is stable as long as $ST - PR > 0$, such that $\lambda_1 < 0$.
A stable homogeneous steady state, with coexistence of both types within each patch exists if and only if $P < S$ and $R < T$, proving that shortwave instability (\emph{e.g.}, Fig.~3) cannot occur in the prisoner's dilemma since it violates these conditions by definition \cite{doebeli2005models}.

\subsection*{Spatial networks and dispersal}
To generate larger networks for simulations (Fig.~\ref{fig:fig4}), we randomly assign coordinates drawn from a uniform distribution $\sim U(0, 1)$ to patches in a 2 dimensional space. Patches are connected if the Euclidean distance between their coordinates falls below a threshold value $h = 0.195$. Simulations were conducted with the Mathematica 12.3.1.0 software.

To define the defector avoidance rule for emigrating from patches, first suppose that from the perspective of a cooperator, interactions occur at random time points, amounting to some effective rate $r$ (\textsl{i.e.,}~a Poisson process). 
Assume further that in each interaction, the player is cheated with some probability $p$. In a sequence of $n$ interactions, we find $n - \alpha + 1$ sub-sequences of $\alpha$ consecutive events, which can be treated as independent trials to very good accuracy. 
Each of the sub-sequences will consist of $\alpha$ cheating events with probability $p^{\alpha}$, and so the rate at which the player experiences $\alpha$ consecutive cheating events and then leaves is $r p^{\alpha}$, explaining the form of Eq.~\ref{eq:gen_m_diff}.

\subsection*{A master stability function approach}
If we start in a homogeneous state, we cannot observe a beneficial effect of any dispersal strategy (\emph{e.g.}, defector avoidance) unless the homogeneous state loses stability, the system departs from the homogeneous state, and spatial patterns begin to form. Such patterns are characterized by an unequal distribution of cooperators and defectors, which can benefit cooperators.  

To explore the stability of the homogeneous state we compute the Jacobian matrix $\bf J$, with a $N M\times N M$ dimension, where $N$ is the number of player types and $M$ is the number of patches in the spatial network. The Jacobian of the spatial system in the compact form can then be expressed as
\begin{linenomath}\begin{equation}
{\bf J} = {\bf I} \otimes {\bf P} - {\bf L} \otimes {\bf C},
\end{equation}\end{linenomath}
where ${\bf I}$ is identity matrix, ${\bf L}$ is Laplacian matrix of the spatial network ($M \times M$), and ${\bf C}$ is the coupling matrix ($N \times N$).
The Laplacian matrix is constructed
by setting $L_{ii} = \sum_j{A_{ij}} $ and subtracting ${\bf A}$, where ${\bf A}$ is the weighted adjacency matrix.
For the case of defector avoidance, the coupling matrix is 
\begin{equation}
\begin{split}
{\bf C} &= \left(\begin{array}{c c} 
    \partial_C E_{C} & \partial_D E_{C} \\
    \partial_C E_{D} & \partial_D E_{D} \\
    \end{array} \right)\\
&= \left(
\begin{array}{cc}
  \left[(1+\alpha)(R-T)+P-S\right] \phi^{1+\alpha} & -\alpha \phi^{1+\alpha} \\
 0 & 1 \\
\end{array}
\right),
\end{split}
\end{equation}
where
\begin{equation}
\phi = \frac{P-S}{R-T+P-S}
\end{equation}

As the matrix $\bf J$ has a block structure, its eigenvectors also have a similar structure \cite{brechtel2018master},
$\bm{w} = \bm{v} \otimes \bm{q}$, where $\bm{v}$ is an $N$-dimensional vector and $\bm{q}$ is an $M$-dimensional vector. Let $\bm{v}$ be an eigenvector of ${\bf L}$ with eigenvalue $\kappa$, such that ${\bf L} \bm{v} = \kappa \bm{v}$.
Also, let $\bm{q}$ be an eigenvector of ${\bf P} - \kappa {\bf C}$ with eigenvalue $\lambda$. Then, $\bm{w}$ is an eigenvector of ${\bf J}$ with eigenvalue $\lambda$ as the following calculation shows:
\begin{eqnarray}
{\bf J} \bm{w} &=& ({\bf I} \otimes {\bf P} - {\bf L} \otimes {\bf C}) \cdot (\bm{v} \otimes \bm{q})\\
&=& {\bf I} \bm{v} \otimes {\bf P} \bm{q} - {\bf L} \bm{v} \otimes {\bf C} \bm{q}\\
&=& \bm{v} \otimes {\bf P} \bm{q} - \kappa \bm{v} \otimes {\bf C} \bm{q}\\
&=& \bm{v} \otimes ({\bf P} - \kappa {\bf C}) \bm{q}\\
&=& \bm{v} \otimes \lambda \bm{q} = \lambda (\bm{v} \otimes \bm{q}) = \lambda \bm{w}
\end{eqnarray}
Since all eigenvectors of ${\bf J}$ can be constructed in this way, the complete spectrum of ${\bf J}$ is then 
\begin{linenomath}\begin{equation}\label{eq:eigvals}
{\rm Ev}({\bf J})=\bigcup_m^M {\rm Ev}( {\bf P} - \kappa_m {\bf C}),
\end{equation}\end{linenomath}
where $\kappa_m$ is the $m$th eigenvalue of ${\bf L}$ \cite{brechtel2018master}. Since every Laplacian eigenvalue $\kappa_i$ generates a set of Jacobian eigenvalues which is independent of the other Laplacian eigenvalues, Eq.~\ref{eq:eigvals} defines a master stability function using only knowledge about the local system ($\bf P$) with some minor modifications to account for spatial processes ($\bf C$). This method therefore permits the fast computation of the leading Jacobian eigenvalue for a given Laplacian eigenvalue.

The resulting function $S(\kappa) = \rm{Re}[\lambda_{max}(\kappa)]$ is then a master stability function for the meta-community. To achieve stability, all eigenvalues of the Jacobian need to have negative real parts, which means only when $\mathrm{Tr}(\bf{J}) < \rm 0$ and $\mathrm{Det}(\bf{J}) > \rm 0$ simultaneously, the steady state can be stable. Stability is lost if any Laplacian eigenvalue falls into a range where the master stability function is positive. This enables us to analyze the stability of the spatial reaction-diffusion system by first computing the spectrum of the Laplacian matrix. 

\section*{Acknowledgments}
\textcolor{mySig}{\small We thank U.~Bhatt, A.M.~Hein, B.T.~Martin, and S.~Munch for helpful discussions. A.K.F. was supported by the Research Associateship Program from the National Research Council of the National Academies of Sciences, Engineering, and Mathematics. F.Z. was supported by the China Scholarship Council–University of Bristol Joint Scholarships Programme. S.A.L. was supported by the Army Research Office Grant W911NF-18-1-0325. T.G. was supported by the Ministry for Science and Culture of Lower Saxony (HIFMB project) and the Volkswagen Foundation (grant number ZN3285).}

\bibliographystyle{vancouver}
\bibliography{pnas-sample}

\end{document}